# Grid Market Directory: A Web Services based Grid Service Publication Directory

Jia Yu, Srikumar Venugopal, and Rajkumar Buyya

**Gri**d Computing and **D**istributed **S**ystems (GRIDS) Laboratory
Department of Computer Science and Software Engineering
The University of Melbourne, Australia
Email: {jiayu, srikumar, raj}@cs.mu.oz.au

## Abstract

As Grids are emerging as the next generation service-oriented computing platforms, they need to support Grid economy that helps in the management of supply and demand for resources and offers an economic incentive for Grid resource providers. To enable this Grid economy, a market-like Grid environment including an infrastructure that supports the publication of services and their discovery is needed. As part of the Gridbus project, we proposed and have developed a *Grid Market Directory* (GMD) that serves as a registry for high-level service publication and discovery in Virtual Organisations.

## 1. Introduction

Computational Grids [1] are emerging as the next-generation computing platform and global cyber-infrastructure for solving large-scale problems in science, engineering and business. They enable the sharing, exchange, discovery, selection and aggregation of geographically distributed, heterogeneous resources—such as computers, data sources, visualization devices and scientific instruments. As the Grid comprises of a wide variety of resources owned by different organizations with different goals, the resource management and quality of service provision in Grid computing environments is a challenging task. Grid economy [9] facilitates the management of supply and demand for resources. Also, it enables the sustained sharing of resources by providing an incentive for Grid Service Providers (GSPs).

It has been envisioned that Grids enable the creation of Virtual Organizations (VOs) [19] and Virtual Enterprises (VEs) [18] or computing marketplaces [20]. A group of participants with a common objective can form a VO. Organizations or businesses or individuals can participate in one or more VOs by sharing some or all of their resources. To realize this vision, Grids need to support diverse infrastructures/services [19] including an infrastructure that allows (a) the creation of one or more VO(s) registries to which participants can register themselves; (b) participants to register themselves as GSPs and publication of resources or application services that they interested in sharing; (c) GSPs to register themselves in one or more VOs and specify the kind of resources/services that they would like to share in VOs of their interest; and (d) the discovery of resources/services and their attributes (e.g., access price and constraints) by higher-level Grid applications or services such as Grid resource brokers. These services are among fundamental requirements necessary for the realisation of Grid economy.

Several Grid economy models drawn from conventional markets have been proposed for organizing the Grid market [9]. They are: commodity, posted price, bargaining, tender/contract and auction models. In Grid economy models, a trusted third party, *Service Publication Directory*, is needed as a central service linking resource providers and consumers. For example, in the commodity model, resource providers publish their services to a Directory, providing service location, service type and service charge price, etc., while resource brokers query the directory and select a suitable service according to the quality-of-service (QoS) requirements (e.g. deadline and budget) of their delegating consumers.

In this paper, we propose a service publication and discovery registry called the *Grid Market Directory* (GMD) that meets the above requirements. The GMD has been developed using emerging web services technologies as part of the Gridbus Project, which is developing technologies that provide end-to-end support for resource allocation based on the QoS requirements of resource providers and consumers.

The rest of this paper is organized as follows. The related work within Grid and Web services communities is presented in Section 2. The detailed system architecture and design issues are described in Section 3. Section 4 describes technologies that are used in the current implementation. A use-case study is presented in Section 5. We conclude in Section 6 with a discussion of current system status and future work.

## 2. Related Work

The two most related/complimentary technologies are Globus MDS [21] and UDDI [15]. The UDDI targeted on providing publication and discovery mechanisms for web services. The Globus MDS, built using LDAP [7] technologies, is designed to provide grid information services such as resource characteristics and status information. We believe that the Gridbus GMD is complimentary to the Globus MDS as it provides higher-level services such as serving a registry of participants in VOs. Additionally, it is designed to enable the notion of Grid economy.

## 3. GMD Architecture and Design

The architecture of the GMD is shown in Figure 1. The key components of the GMD are:

- GMD Portal Manager (GPM) facilitates service publication, management and browsing. It allows service providers and consumers to use a web browser as a simple graphical client to access the GMD.

- GMD Query Web Service (GQWS) enables applications (e.g. resource broker) to query to search the GMD, and find a suitable service to meet the job execution requirement (e.g. budget) and the price of a particular service for cost calculation, resource allocation scheduling, etc.

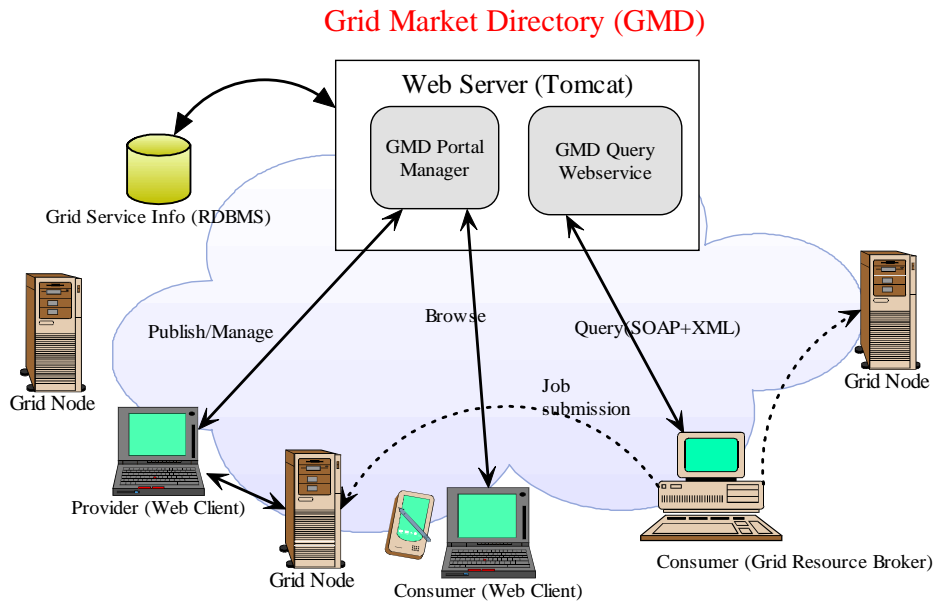

**Figure 1**: Gridbus GMD Architecture

Both the components receive client requests through a HTTP server. Additionally, a database (GMD repository) is configured for recording the information of Grid services and service providers.

### 3.1 GMD Portal Manager

The architecture of the GMD Portal Manager (GPM) is shown in Figure 2. The GPM provides three different access interfaces: service browsing, provider administration and service management.

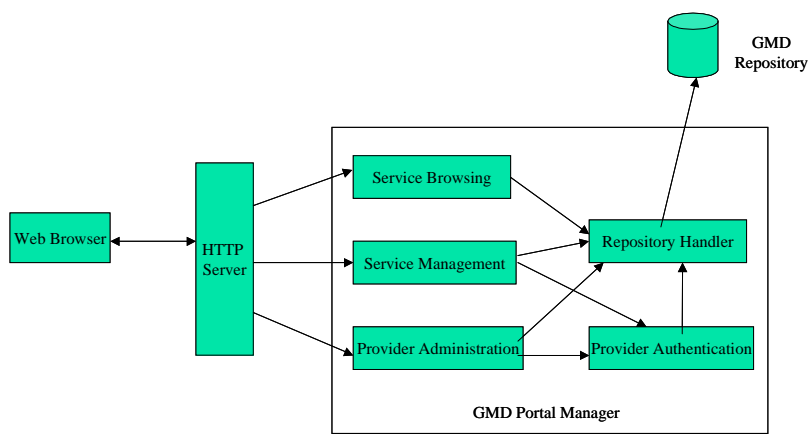

**Figure 2**: GMD Portal Manager Architecture

**Service browsing**
The GPM allows users to browse all registered services and also browse services offered by a specific provider. Additionally, services in the GMD are categorized by service type, such as Earthquake Engineering, Molecular Docking and CPU Service, so that users can browse services for a particular application area. For instance, the high-energy physics community can browse services related to its area along with their access costs.

**Provider administration**
The provider administration module is responsible for account management including provider registration and removal. Provider's account information is acquired at the time of registration. This includes the provider's name, login name, password, contact address and some additional information.

**Service management**
The service management module enables the registered providers to maintain their services in the GMD. A service management page is dynamically generated for each registered provider, through which the provider can add, update and remove services. Basic service attributes include: service name, service type, hardware price (/CPU-sec), software price (/Application Operation), node host name and application path.

In addition, security issues are also addressed in the GPM. A login authentication mechanism for identifying registered providers is employed in the service management and provider administration. In the service management, service modification operations are also authenticated before being committed to the repository.

### 3.2 GMD Query Web Service

The GMD also provides web services [11] for application querying. The GMD Query Web Service (GQWS) is built using SOAP (Simple Object Access Protocol) [12]. The main benefit from using SOAP is multi-language supporting. Since there is support for the SOAP protocol in many programming language, for example, C++, Java and VB, application developers have flexibility of selecting the programming environment of their choice.

The six basic SOAP invocation methods supported by the GQWS are listed below:

1. QueryService() - return a list of all services.

2. QueryServiceByType(serviceType) – return a list of services offered by a certain service type.
3. QueryServiceByHost(hostName) – return the service information associated with a certain host name.
4. QueryServiceByProvider(providerName) – return a list of services provided by a certain provider.
5. QueryServiceContact(serviceType) – return a list with only name and address information of services provided by a specified service type.
6. QueryPrice(serviceName) – return service price for a specified service.

The GQWS consists of two modules: *Query Processor* and *Repository Handler*. The interaction between the GQWS and its client is illustrated in Figure 3. The GQWS communicates with its client using XML [10] formatted message. The XML-formatted query message is encapsulated within a SOAP message, which is transferred by HTTP between the web server and a GMD client. The SOAP engine acquires the query message and forwards it to the GQWS. The *Query Processor* handles query message parsing and takes appropriate actions based on the content of the message, while *Repository Handler* is responsible for retrieving data from database. The XML-formatted response message is finally constructed by *Query Processor* and sent back to the client.

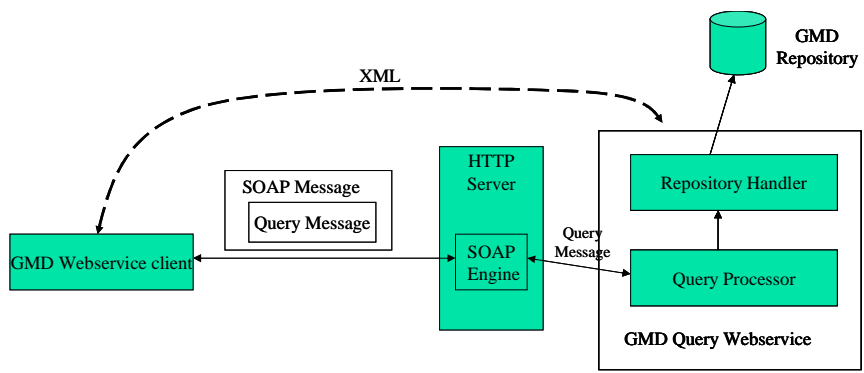

**Figure 3:** Interaction between the GQWS and its client

The format of the query message for indicating a certain service type (e.g. CPU service) is:

```
<query_service>
        <service_type>CPU service</service_type>
</query_service>
```

This XML structured query message can be extended easily. The example below illustrates a query message with two constraints: service type and service provider.

```
<query_service>
        <service_type>…</service_type>
        <provider_name>…</provider_name>
</query_service>
```

If the query for services of certain type is successful, the *Query Processor* responds with a message whose format is given below:

```
<?xml version="1.0" encoding="UTF-8"?>
<service-details type="…" status="ok">
        <service>
                <name>…</name>
                <provider>…</provider>
                <price>
```

```
                        <hardware>…</hardware>
                        <software>…</software>
                </price>
                <address>…</address>
                <description>…</description>
        </service>
        <service>
                        .
                        .
                        .
        </service>
        <service>
                        .
                        .
                        .
        </service>
</service-details>
```

The attribute *type* indicates the service type of the services listed in the response message. The attribute *status* indicates the processing status of the query. If the query fails, the format of the response message appears as follows:

```
<?xml version="1.0" encoding="UTF-8"?>
<service-details type="…" status="error">
        <reason>…</reason>
</service-details>
```

The detailed specification of XML elements used in the GMD can be found in [2].

**GQWS Client API Architecture**

In addition to the standard SOAP client infrastructure, the GMD clients need XML message building and parsing capability depending on the query and response specification. The GMD supports Java client APIs that hide the specifics of SOAP/XML and aids in the rapid development of GMD applications. Thus, the developers will be able to utilize the GQWS into their applications easily without the knowledge of low-level details of SOAP and XML usage. Figure 4 presents the architecture of the GMD client APIs and their interaction with the GQWS.

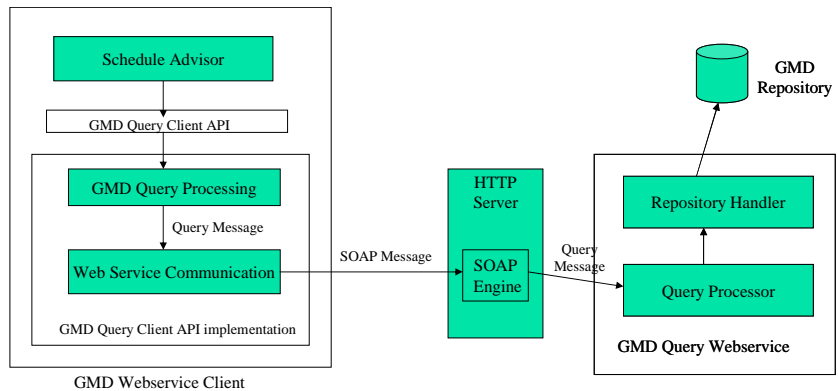

**Figure 4:** GMD Query Client API Architecture

## 4. Implementation

We have constructed a Grid Market Directory, which implements the architecture described above, under Windows and Solaris platform. The GMD was implemented by using various Internet/Web technologies. They include Jakarta Tomcat[14], Apache SOAP[3], JDOM [6] and MySQL[8]. Jakarta Tomcat is used for providing HTTP service and Apache SOAP is used for the SOAP implementation. JDOM is used for XML message building and parsing. MySQL is chosen for persistence storage of the service information and provider profile.

The presentation view of the Portal Manager is implemented by using Java Server Pages (JSP) [4]. The main benefit of using JSP is that it allows the instantiation of Java objects within HTML. Thus, the dynamic display of the data retrieved from the GMD database using JDBC [23] can be compiled into HTML. Two pages generated by the Portal Manager are shown in Figure 5.

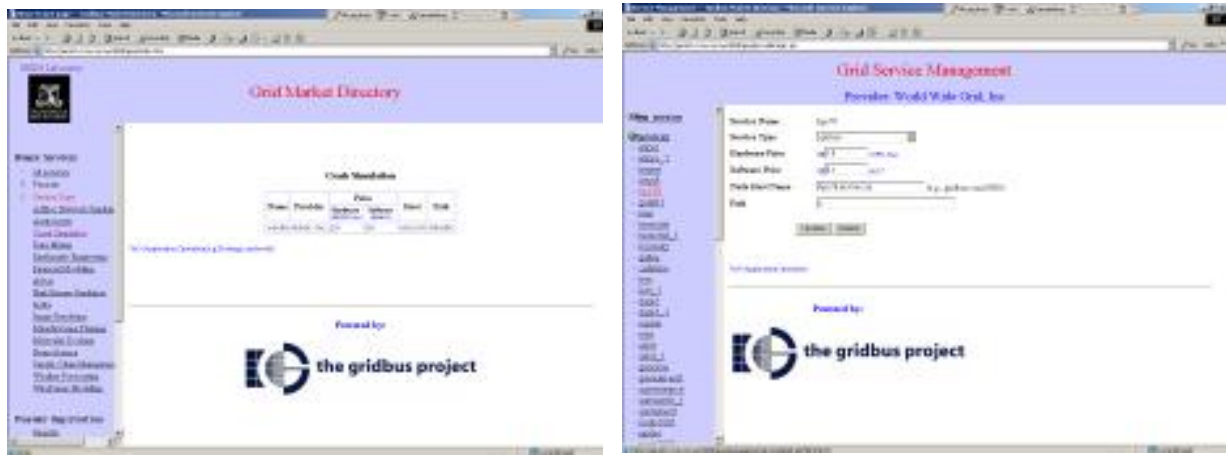

**Figure 5**: Portal Manager Demo pages from left to right (a) service browsing by type
*Crash Simulation* page (b) service modification page for *World Wide Grid, Inc.*

Service information modification supported by the GMD website is protected by the login name and password. Java servlet HttpSession API[5] is used for the GMD security implementation. As shown in figure 6, after logging a name attribute is set to the HttpSession object of corresponding request session. When the user logs out, the name attribute is deleted from the object. Therefore, it can be known whether the user of the request has logged in by checking the name attribute. If the request is illegal (unable to authenticate), the GMD will not permit access to the GSP service management page.

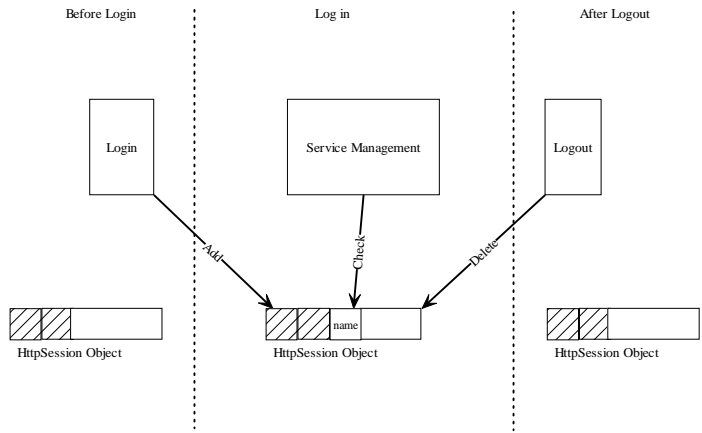

**Figure 6:** Security Mechanism of GMD Portal Manager

## 5. Use Case Study

We have participated in the Global Grid Testbed Collaboration [22] that setup a world-wide distributed Grid testbed and demonstrated capabilities of the state-of-the-art Grid technologies and applications at the SC 2002 conference [13] held in Baltimore, USA. The testbed had resources contributed by participating organizations from all over the world. We had access to most resources in the testbed as we were demonstrating one of the four Grid applications as part of this collaboration. As this collaboration fits to the notion of virtual organisation (VO), we used the GMD infrastructure to create a VO registry and added participants as Grid service providers along with their contributed resources/services and their attributes. The GMD used in the creation of this VO registry was hosted on a (Sun Solaris) server located at the University of Melbourne, Australia (see Figure 7). In this VO, the GMD services have been used for the following purposes:

- To register the testbed participants as Grid service providers
- To publish resources/services contributed by each participant and their attributes.
- To enable GMD applications/tools to discover resources or application services at runtime.

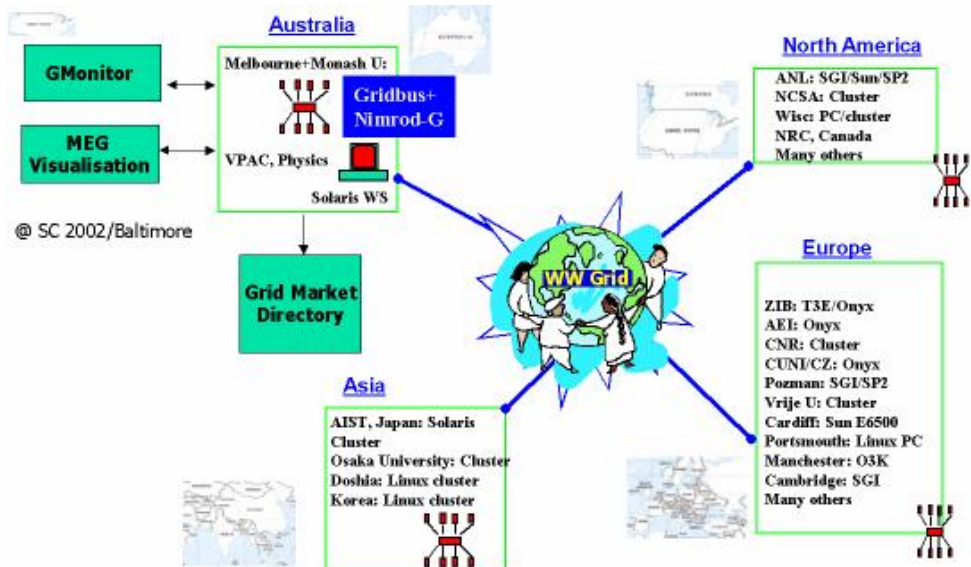

**Figure 7:** GMD Usage in the HPC Challenge Demo @ SC 2002.

We have demonstrated distributed execution of brain activity analysis application on Global Grids using the Nimrod-G broker [16] with the Gridbus scheduler [17]. The Gridbus scheduler, implemented as a plug-in scheduler for Nimrod-G, supports cost-based scheduling and applications. The default Nimrod-G scheduler takes service price from a user configuration file, while the Gridbus scheduler identifies resources/services along with their attributes such as access-price by querying the GMD. Additionally, the Gridbus scheduler has been enhanced to support resource allocation based on A*pplication Operation(AO)* service price in addition to traditional CPU cost. The results and scheduling experiments that used the GMD service are reported in [17].

## 6. Conclusion and Future Work

As the Grid economy-based management of distributed resources helps in management of the supply and demand for resources and offers an economic incentive for sustaining resource sharing and aggregation, an infrastructure that supports the publication of services and their discovery in a market-like environment is needed. In this paper, we proposed and have developed an emerging web services based Grid service publication and discovery infrastructure called the Grid Market Directory (GMD). The GMD consists of two key components: the Portal Manager and Query Web Service. The GMD Portal Manager is responsible for provider registration, service publication and management, and service browsing. All these tasks are accomplished by using a standard web browser. The GMD Query Web Service provides services so that clients such as resource brokers can query the GMD and obtain the information of resources to discover those that satisfy the user QoS requirements

The GMD has been implemented as part of the *open source* Gridbus project. The GMD makes use of commodity Java libraries including Java Server Page, Apache SOAP, Java Servlet API, JDBC API and JDOM. An open-source servlet container, Tomcat, is used to provide HTTP services and JDBC supporting database, MySQL, is used to provide the GMD repository services.

The current version of the GMD supports commodity market model and its services are being used by clients such as economy-based grid schedulers. Future work will enhance the GMD by supporting other economy models such as tender/contract and auction models. Furthermore, efforts are currently underway to make it OGSA-compliant [25] and to enable the publication of OGSA-compliant higher-level (application) services.

### Availability

The GMD software with the source code can be downloaded from the Gridbus project website:

> http://www.gridbus.org/gmd/

### Acknowledgements

We would like to thank Steve Melnikoff and Anthony Sulistio for their comments on the paper.